\documentclass[12pt,twoside]{article}
\usepackage{hyperref}
\usepackage{adjustbox}
\usepackage{geometry}
\usepackage{xcolor}
\usepackage{fancyhdr}
\usepackage{amsmath}

\usepackage[labelfont=bf]{caption}
\usepackage{natbib}
\usepackage{graphicx}
\usepackage{float}
\usepackage{booktabs}
\usepackage{tabularx}
\usepackage{titling}
\usepackage{adjustbox}
\usepackage{titlesec}
\titleformat{\section}{\normalfont\bfseries\filcenter}{}{0pt}{}
\titleformat{\subsection}{\normalfont\bfseries\filcenter}{}{0pt}{\itshape}
\titleformat{\subsubsection}{\normalfont\bfseries\filcenter}{}{0pt}{\itshape}
\titlespacing*{\section}
  {0pt}{-.1\baselineskip}{-.1\baselineskip}  
\titlespacing*{\subsection}
   {0pt}{-.1\baselineskip}{-.1\baselineskip}

\date{}
\providecommand{\keywords}[1]
{
   \small	
  \textit{\hspace{-1em} Keywords: } #1
}

\fancypagestyle{firstpage}{
  \fancyhf{}
  \fancyfoot[L]{\color{gray} \footnotesize Copyright \copyright 2022. Licensed under the Creative Commons Attribution Non-commercial No Derivatives (by-nc-nd).}%
  \lhead{\color{gray} \footnotesize Preprint.}
\rhead{\color{gray} \footnotesize Carebot s.r.o.}
}
\title{\normalsize \textbf{TOWARDS CLINICAL PRACTICE: DESIGN AND IMPLEMENTATION OF CONVOLUTIONAL NEURAL NETWORK-BASED ASSISTIVE DIAGNOSIS SYSTEM FOR COVID-19 CASE DETECTION FROM CHEST X-RAY IMAGES}} 
\author{\normalsize Daniel Kvak, MSc$^{\ast}$ • Marian Bendik • Anna Chromcova, MD \\ \thanks{$^{\ast}$Corresponding author: daniel.kvak@carebot.com}  \thanks{Date submitted: 2022-03-18} \thanks{} \thanks{This is an open access article distributed under the Creative Commons Attribution License which permits unrestricted use, distribution, and reproduction in any medium, provided the original work is properly cited.}} 
\usepackage{lipsum}
\renewcommand\footnotemark{}
\begin{document}
\maketitle
\thispagestyle{firstpage}
\vspace{-5em}
\begin{abstract}
  \noindent  One of the critical tools for early detection and subsequent evaluation of the incidence of lung diseases is chest radiography. This study presents a real-world implementation of a convolutional neural network (CNN) based Carebot Covid app to detect COVID-19 from chest X-ray (CXR) images. Our proposed model takes the form of a simple and intuitive application. Used CNN can be deployed as a STOW-RS prediction endpoint for direct implementation into DICOM viewers. The results of this study show that the deep learning model based on DenseNet and ResNet architecture can detect SARS-CoV-2 from CXR images with precision of 0.981, recall of 0.962 and AP of 0.993.
\end{abstract}

\keywords{\textit{computer-aided detection, convolutional neural network, COVID-19, deep learning, image classification.} \vspace{8ex}}
\section{1 Introduction}
COVID-19 disease causes severe acute respiratory syndrome 2 (SARS-CoV-2), which was originally discovered in late 2019. On January 30, 2020, the outbreak was declared a Public Health Emergency of International Concern and subsequently, on March 11, 2020, WHO declared COVID-19 a pandemic. \citep{wu2020outbreak} Once infected, a patient with COVID-19 may develop various symptoms and signs of infection, which include fever, cough, and respiratory illness. In severe cases, infection can cause pneumonia, respiratory distress, multi-organ failure and death. \citep{cascella2022features, shereen2020covid, rothan2020epidemiology}

At a time when the speed and reliability of results, especially for COVID-19 positive patients, is important, the development of applications that would facilitate the work of untrained staff involved in the evaluation is also crucial. Chest radiography is an important tool for early detection and subsequent verification of lung diseases. \citep{qin2018computer, speets2006chest} As such, radiographic testing can be performed more quickly and has greater accessibility due to the prevalence of chest radiological imaging systems in modern healthcare systems and the availability of portable units, making them a suitable adjunct to RT-PCR testing, particularly as CXR imaging is often performed as part of the standard procedure for patients with respiratory difficulties. \citep{MARTINEZCHAMORRO202156, kvak2021automatic} However, in the current pandemic situation of COVID-19, we encounter a lack of radiologists and trained staff to analyze the vast number of images taken. \citep{cavallo2020economic}

Many researchers have published a series of preprints demonstrating computer-aided detection (CADe)-based approaches to detect COVID-19 from chest radiographs. \citep{apostolopoulos2020covid, 10.3389/fmed.2020.00550, 9591581} These approaches have achieved promising results on a small dataset but are by no means production-ready solutions. The aim of our research is to present a real-world CNN-based approach that would eliminate the time and resources needed to develop new technologies and related algorithms. \citep{meenatchi2018survey}

\begin{table}
\centering
  \begin{adjustbox}{width=1\textwidth}
\begin{tabular}{lllll}
\hline
Work & Dataset                                                                                                                    & Model                                                              & Accuracy & F1 Score \\ \hline
\citep{apostolopoulos2020extracting}    & \begin{tabular}[c]{@{}l@{}}455 COVID-19, 2109 Non-COVID\\ Images\end{tabular}                                              & MobileNet V2                                                              & 99.18\%  & -        \\
\citep{MAHMUD2020103869}    & \begin{tabular}[c]{@{}l@{}}305 COVID-19, 1888 Normal, 3085\\ Bacterial Pneumonia, 1798 Viral\\ Pneumonia\end{tabular}      & \begin{tabular}[c]{@{}l@{}}Stacked MultiResolution\\ CovXNet\end{tabular} & 97.4\%   & 0.971    \\
\citep{ozturk2020automated}    & \begin{tabular}[c]{@{}l@{}}127 COVID-19, 500 Normal, 500\\ Pneumonia Images\end{tabular}                                   & \begin{tabular}[c]{@{}l@{}}DarkCovidNet\\ \end{tabular}              & 98.08\%  & 0.965    \\
\citep{el2021using}    & \begin{tabular}[c]{@{}l@{}}231 COVID-19, 1583 Normal, 2780\\ Bacterial Pneumonia, 1493 Viral Pneumonia Images\end{tabular} & \begin{tabular}[c]{@{}l@{}}Inception\\ ResNetV2\end{tabular}              & 92.18\%    & 0.921    \\
\citep{KHAN2020105581}    & \begin{tabular}[c]{@{}l@{}}284 COVID-19, 310 Normal, 330\\ Bacterial Pneumonia, 327 Viral\\ Pneumonia Images\end{tabular}  & CoroNet                                                                   & 99\%     & 0.98     \\ \hline
\end{tabular}
  \end{adjustbox}
  \caption{\label{tab:table-name}Comparison with state-of-the-art methods.}
\end{table}

\newpage
\section{2 Radiology perspective}
The COVID-19 disease can be identified on a conventional chest radiograph based on several typical patterns. The two most common patterns are ground-glass opacities and lung consolidation. However, the ground-glass opacities observed on a chest CT can be difficult to detect on a chest radiograph, it is often accompanied by a reticular opacities region which is more easily appreciable on a standard CXR. \citep{cozzi2021ground} Lung consolidations in COVID-19 (and other viral pneumonia) can be often found multifocally, the distribution is usually bilateral and includes lower lobes. \citep{jacobi2020portable} One of the most specific features of COVID-19 on a CXR is a peripheral and posterior distribution of air-space opacities. In more serious stages of the disease when patients are typically hypoxic, diffuse air-space opacities covering the majority of lung parenchyma can be found and the CXR pattern can be similar to acute respiratory distress syndrome (ARDS). The rare or less common findings in COVID-19 are pleural effusions, pneumothorax, and lung cavitation. \citep{jacobi2020portable}

During the first four days of symptomatic COVID-19 disease, there can be a normal finding on the chest radiograph and chest CT scan. Later on, when there are present the signs of COVID-19 on CXR, the images can show a high similarity to those of several types of viral pneumonia and other inflammatory lung diseases. Therefore, it is difficult for medical doctors to distinguish COVID-19 infections from other viral pneumonia using only a chest X-ray. However, the conventional chest radiograph as well as the chest CT scan were found to be an important diagnostic tool in addition to PCR test for their higher sensitivity, availability, speed, and possible prediction of severity of the disease. \citep{tahir2021covid}

\section{3 Proposed model architecture}
In the recent past, deep learning has been very successful in a variety of visual tasks. Deep learning-based models have revolutionized CADe by accurately analyzing, identifying and classifying patterns in medical images. \citep{yamashita2018convolutional} In the past, deep learning has had success in mammography image classification. \citep{shen2019deep} The success of machine learning-based algorithms in CADe and the rapid growth of COVID-19 cases have necessitated the need for an automatic detection and diagnosis system based on artificial intelligence. Recently, many researchers have proposed the use of CNN-based CADe models to detect COVID-19 from CXR. \citep{apostolopoulos2020extracting, MAHMUD2020103869, ozturk2020automated, el2021using, KHAN2020105581}

In recent years, the advent of a new generation of GPUs and the deployment of cloud environments such as Microsoft Azure and Amazon AWS have kick-started a second wave of innovation in CADe. \citep{kagadis2013cloud} Unlike earlier approaches which utilized handcrafted features, the CNN-based models are trained to identify relevant biomarkers on a large dataset without explicit input from researchers. 

\subsection{3.1 Convolutional neural network}
For image recognition and classification tasks, various CNN architectures have proven to be widely used. The basic idea is that neurons in the visual cortex process images into increasingly complex shapes. \citep{NIPS2012_c399862d} The image is first segmented at the boundaries of edges using a light/dark interface, then fused into simple shapes, and finally fused into recognizable complex features in subsequent layers. \citep{NIPS2012_c399862d} CNN tries to mimic this idea using multiple layers of artificial neurons. The standard architecture includes several convolutional layers that segment the image into small pieces that can be easily processed. \citep{8308186}

\begin{table}[]
\centering
\begin{adjustbox}{width=0.8\textwidth}
\begin{tabular}{@{}lllll@{}}
\toprule
\textbf{Model} & \textbf{Layers} & \textbf{Parametres} & \textbf{Input matrix} & \textbf{Output activation} \\ \midrule
ResNet50V2 \citep{he2016deep}     & 50              & 25.6M               & 224, 224, 3           & softmax                    \\
DenseNet121 \citep{huang2017densely}     & 121             & 8M                  & 224, 224, 3           & softmax                    \\ \bottomrule
\end{tabular}
\end{adjustbox}
\caption{\label{tab:table-name}Architecture of the proposed models implemented in Carebot Covid app.}
\end{table}

The outputs from these layers are aggregated into layers to reduce the size of the data and reduce noise. The sequential layers feed into a neural network, which then produces a probabilistic heat map that describes the probability of whether the image contains the desired target. \citep{NIPS2012_c399862d} The advantage of this system is that it can be trained to find any feature in an image without the designer having to describe specific features. \citep{yamashita2018convolutional} The convolutional layer is an important part of the deep learning neural network that extracts features from the input images. This avoids a major limitation of previous hand-crafted algorithms, which required hard-coding specific identifying features into the software. 

\begin{figure}[H]
\includegraphics[width=1\textwidth]{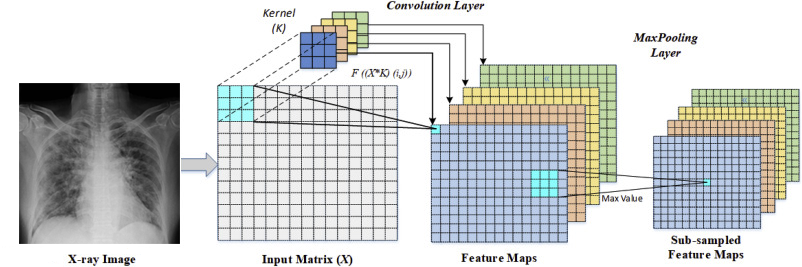}
\centering
\caption{\textbf\scriptsize{Schematic representation of convolution and max-pooling layer operations inside a CNN.}}
\end{figure}

\subsection{3.2 DICOMweb implementation}
Current paradigms for the use of medical image resources call for vendor-neutral archives, accessible through standard interfaces and with support for multiple repositories. The communication processes, data format, storage, querying, retrieval, visualisation and printing of this medical image information are specified by the international standard Digital Imaging and Communications in Medicine (DICOM), which is the main standardisation effort in this field. \citep{lebre2020cloud} Patient information is bundled in one or more standard files that contain metadata related to the patient, study, or report in addition to image pixels. A typical infrastructure consists of one or more archives, acquisition methods (data production units), distribution mechanisms, and visualization devices. \citep{lebre2020cloud}

The DICOMweb STOW-RS (STore Over the Web) service is the basis of the IHE WIC integration profile. \citep{clunie2016technical} WIC seeks to simplify the sender's task by allowing not only the sending of DICOM objects, but also the transmission of non-DICOM format images or videos along with minimal DICOM metadata in XML or JSON encoding, and by shifting the burden of filling in the DICOM metadata describing the pixel data to the server. \citep{6943849}

\begin{figure}[H]
\includegraphics[width=1\textwidth]{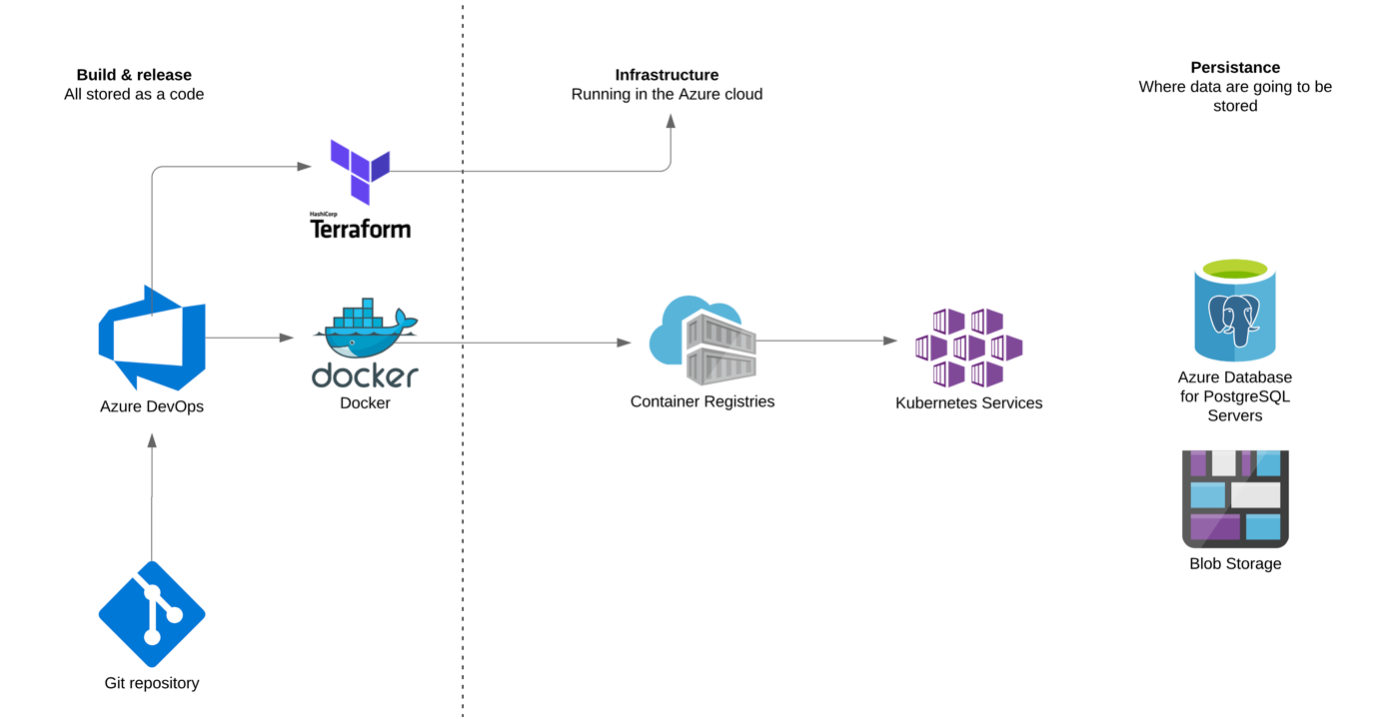}
\centering
\caption{\textbf\scriptsize{Production environment high-level architecture of Carebot Covid app.}}
\end{figure}

STOW-RS has been defined to reduce the burden for implementing lightweight dispatch applications, for example on mobile devices. \citep{genereaux2018dicomweb} This allows the sender to use an HTTP POST operation to send one or more DICOM objects, either encoded as regular DICOM PS3.10 binary files, or as separate XML or JavaScript Object Notation (JSON) metadata and separate image or video pixel data in various Internet media file formats. \citep{6943849}

Communication in the opposite direction, i.e. from the prediction server back to the user, is handled by the DICOMweb WADO-RS (Web Access to Dicom Persistent Objects) middleware. The endpoint is integrated into the Picture Archiving and Communication Systems (PACS) web system and provides services for a specific PACS system. It can be deployed in a hospital to provide the WADO service to healthcare professionals and used in regional PACS to transfer medical images and messages. \citep{liu2015smartwado} The use of WADO-RS allows us to build a DICOM Structured Report object that contains the requested CADe application predictions. 

\subsection{3.3 Out-of-distribution detection}
CNNs are often trained with a closed-world assumption, i.e., the distribution of the test data is assumed to be similar to the distribution of the training data. \citep{yamashita2018convolutional} However, when deployed in real-world, this assumption does not hold, leading to significant performance degradation. Although this performance degradation is acceptable for applications such as product recommendation, it is dangerous to use such systems in intolerant domains such as CADe because they can cause serious accidents. \citep{firmino2016computer} The proposed Carebot Covid app generalizes to out-of-distribution (OOD) examples whenever possible, flagging those that are beyond its capabilities and seeking human intervention. \citep{hendrycks2016baseline}

\begin{figure}[H]
\centering
\includegraphics[width=0.7\textwidth]{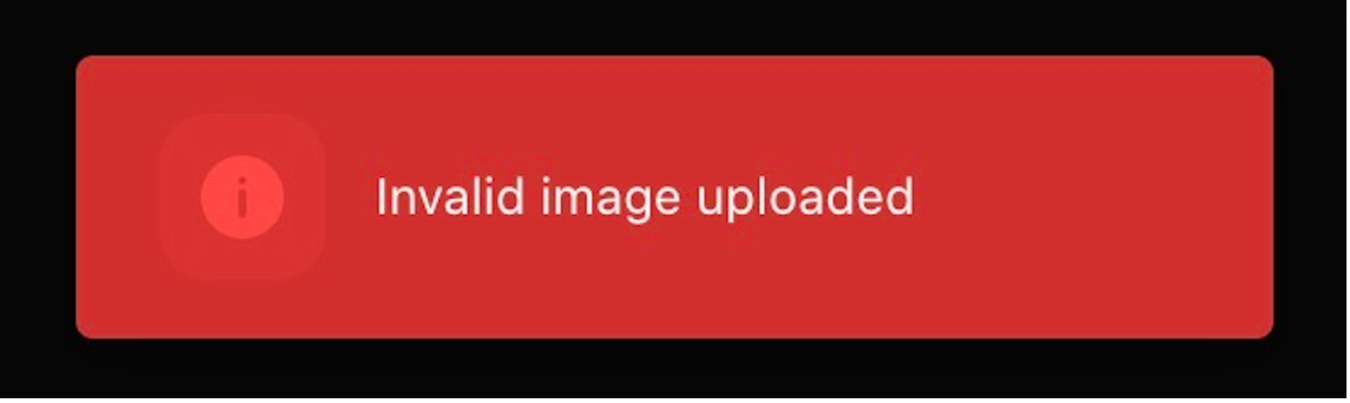}
\caption{\textbf\scriptsize{While similar solutions try to predict the diagnosis even from the inappropriate image, Carebot Covid app successfully detects that it is not a CXR image and discards it as invalid.}}
\end{figure}

One of the difficult tasks is knowing when not to make a prediction. If you ask a radiologist to diagnose an image of something that is not his specialty, he should not provide a diagnosis. For CNN, its specialization is the classifier domain, which is defined by its training distribution. \citep{hermann2020origins} We cannot evaluate the accuracy of the model on these examples, so we choose not to process them, hence only images that are similar to the images in the training distribution can be processed. A CNN-based binary classifier method was selected for model design and calibration, which was trained on a subset of the original dataset and the Tiny ImageNet dataset. \citep{Le2015TinyIV} An OOD model was built on this annotated dataset to predict whether new examples belong to the positive or negative class. \citep{hendrycks2016baseline}

\section{4 Subjects and dataset}
Selected images were obtained from publicly available datasets. The combination of different datasets increases the confidence in the developed identification models, and it also increased the size of the dataset, which is a problem in most of the related literature. \citep{apostolopoulos2020extracting, MAHMUD2020103869, ozturk2020automated, el2021using, KHAN2020105581}

Thus, the full train and test dataset contains 21,905 unique CXRs, of which 3,987 represent CXR images of patients with a positive COVID-19 test, 7,650 CXR images show patients with findings unrelated to COVID-19 infection (pulmonary lesions, fibrosis, pneumothorax etc.), and 10,268 CXR images record patients with no or negligible pathological findings. CXR were taken in posteroanterior (PA) or anteroposterior (AP) projection in patients who were unable to stand. All images in AP projection were taken using portable X-ray machines with patients in supine or sitting position. \citep{zhang2021development}

To create the dataset, we combined and modified several different publicly available datasets. Examples of CXR images from the dataset used are shown in Figure 1 and illustrate the diversity of patient cases (including age, sex, stage of infection, or imaging projection) in the dataset. Patients younger than 15 years were excluded from the dataset, as well as images of poor quality or incorrect projection.

\begin{figure}[H]
\includegraphics[width=1\textwidth]{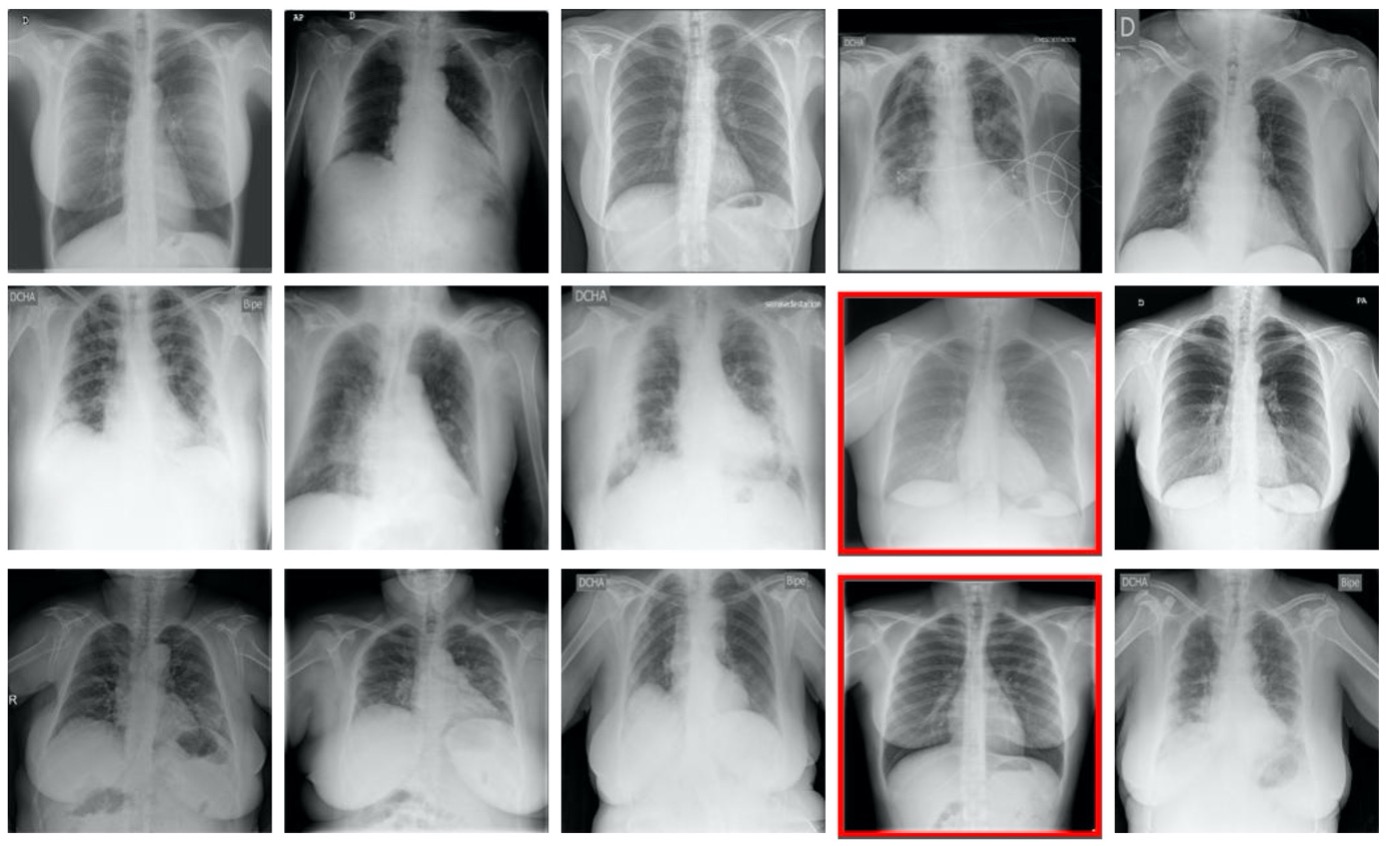}
\caption{\textbf\scriptsize{CXR test images during neural network training. Marked red are those for which CNN incorrectly predicted the class.}}
\end{figure}

\subsection{4.1 Data augmentation}
Data augmentation increases the size of the input training data along with the regularization of the model, thus improving the generalization of the training model. \citep{mikolajczyk2018data} It also helps to create new train examples by randomly applying different transformations to the available dataset to reflect the noisiness of real-world data. \citep{shorten2019survey, elgendi2021effectiveness} In our study, we used transformations involving vertical flipping of training images, random rotations, modifications in lighting conditions, zoom, saturation, and JPEG encoding noise. The training data was augmented using five randomly selected variations, the extension of test dataset was not investigated.

\section{5 Classifier performance}
An F1 Score becomes a critical evaluation tool to determine False Positive and False Negative rates yielded through a discriminating threshold in a similar situation with unbalanced dataset samples. \citep{sokolova2006beyond} The classification performance of the Carebot Covid app model for multi-class problem was evaluated for each component and the average classification performance of the model was calculated. The following table includes the accuracy, precision, recall, and F1 Score, calculated based on the following equations below:

\begin{equation} \label{eqn1}
Accuracy = \frac{TP+TN}{TP+TN+FP+FN}
\end{equation}
\begin{equation} \label{eqn2} 	 
Precision = \frac{TP}{TP+FP}
\end{equation}
\begin{equation} \label{eqn3} 
Recall = \frac{TP}{TP+FN}
\end{equation}
\begin{equation} \label{eqn4} 
F1 Score = \frac{2*Precision*Recall}{Precision+Recall} = \frac{2*TP}{2*TP+FP+FN}
\end{equation}

For specific experiments and given that there is a class imbalance problem, the most reliable metric is the model average accuracy metric, while given that this accuracy is high, the second most important metric is the recall metric for individual classes. \citep{japkowicz2002class} This is due to the importance of correctly identifying true cases that are not COVID-19 (True Negatives). AP (Average Precision) summarizes a precision-recall curve as the weighted mean of precisions achieved at each threshold \citep{yilmaz2006estimating}, with the increase in recall from the previous threshold used as the weight:

\begin{equation} \label{eqn5}
\text{AP} = \sum_n (R_n - R_{n-1}) P_n
\end{equation}
\newpage

\begin{table}

\centering
\begin{tabular}{llllll}
\hline
Class                  & \multicolumn{1}{l}{Image count} & \multicolumn{1}{l}{Precision} & \multicolumn{1}{l}{Recall} & \multicolumn{1}{l}{F1 Score} & \multicolumn{1}{l}{AP} \\ \hline
\textit{Model Average} & 21,905                          & 0.952                         & 0.950                      & 0.951                        & 0.985                  \\
COVID-19               & 3,987                           & 0.981                         & 0.962                      & 0.971                        & 0.993                  \\
Non-COVID-19           & 7,650                           & 0.952                         & 0.922                      & 0.937                        & 0.980                  \\
No Finding             & 10,268                          & 0.941                         & 0.967                      & 0.954                        & 0.984                 
\end{tabular}
\caption{\label{tab:table-name}Averaged test results of Carebot Covid app for multi-class classification after cross validation.}
\end{table}

\begin{figure}[H]
\centering
\includegraphics[width=1\textwidth]{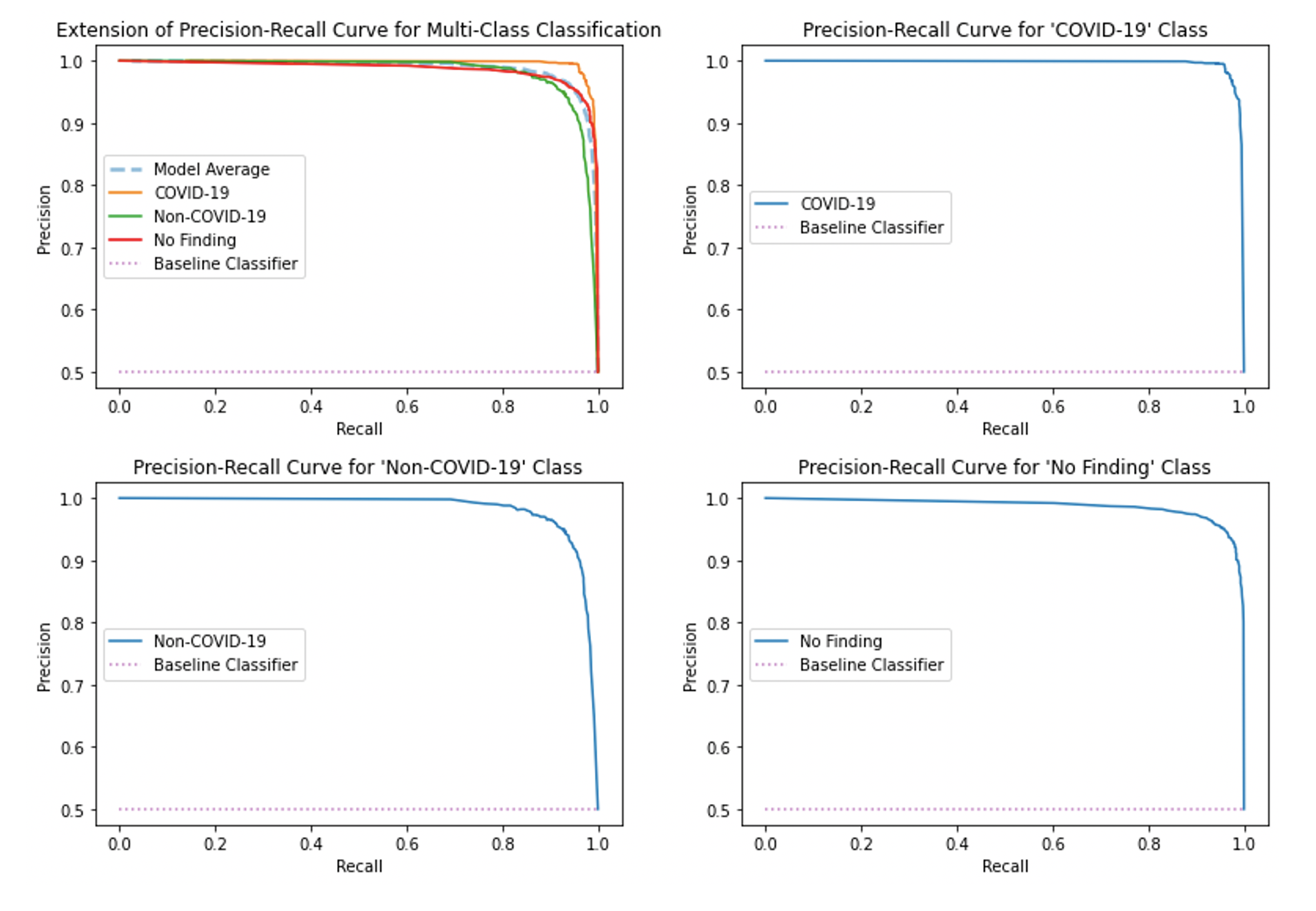}
\caption{\textbf\scriptsize{Evaluation of Carebot Covid app using precision-recall curve for each individual class.}}
\end{figure}
\newpage

The precision-recall curve shows the trade-off between precision and recall for different thresholds. \citep{buckland1994relationship} A high area under the curve represents both high recall and high precision, with high precision associated with low False Positive cases and high recall associated with low False Negative cases. \citep{boyd2013area} Shown baseline classifier is defined as a classifier that cannot distinguish between classes and would predict a random class or a same class in all cases. \citep{brownlee2018use}

\section{6 Conclusion and future work}
In many of the previous studies analyzed, it has been shown that having sufficient clinically annotated data is essential for training more complex CADe applications. \citep{lebre2020cloud, 10.3389/fmed.2020.00550, 9591581} The results of this study showed that the amount of data is also a determining factor for the accuracy of the CNN-based model approach used in the case study. We found that the Carebot Covid app using a deep learning-based model can detect SARS-CoV-2 from CXR images with an accuracy of 0.981, recall of 0.962 and AP of 0.993. In the case of a deep learning-based model, it can be used to detect SARS-CoV-2 from CXR images.

\begin{figure}[H]
\centering
\includegraphics[width=1\textwidth]{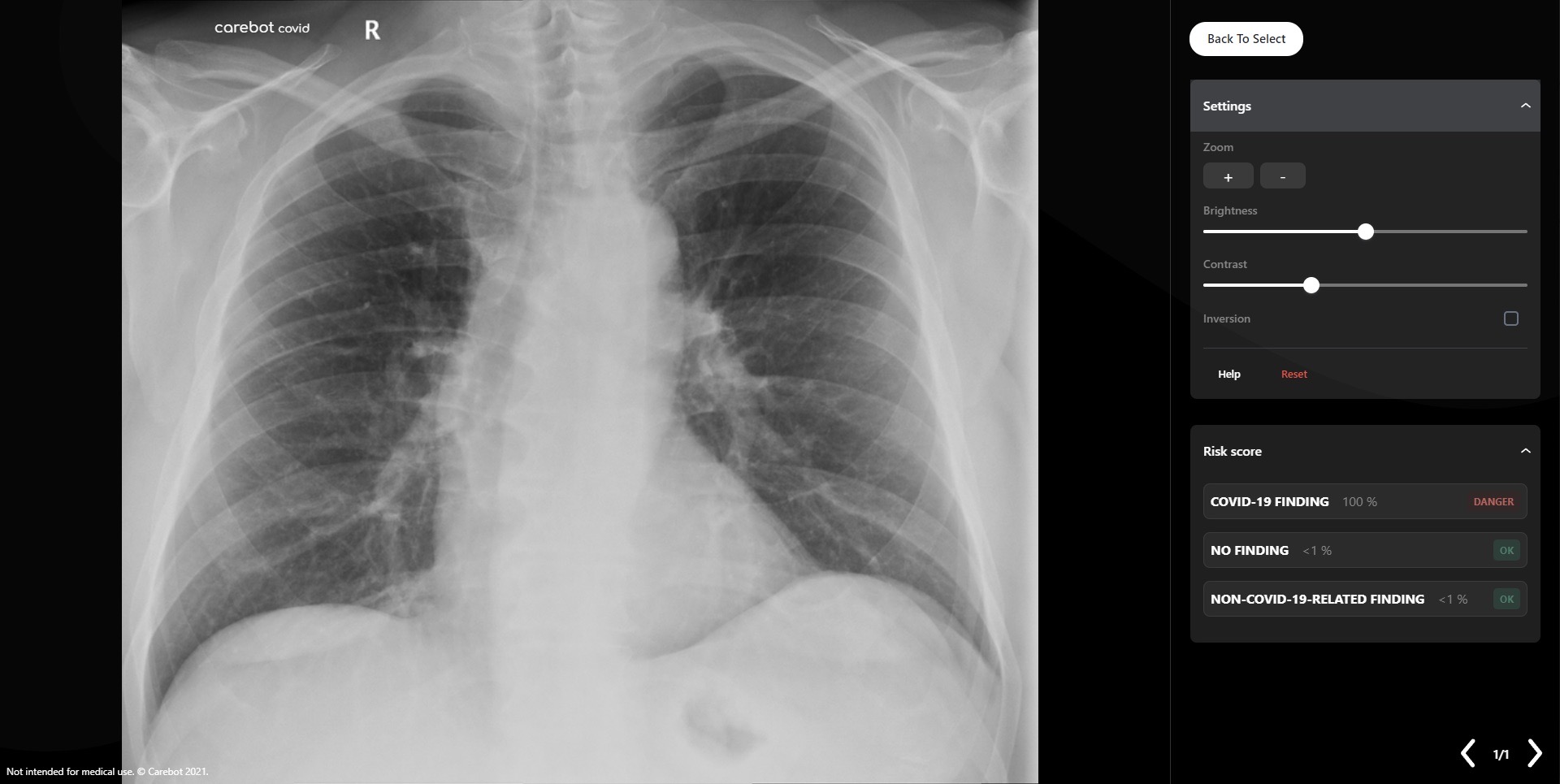}
\caption{\textbf\scriptsize{Simplified GUI of Carebot Covid app.}}
\end{figure}

However, data quantity is not the only element that can be central to an CADe system; the machine learning algorithm used and its parameters, as well as the characteristics of the added data, have a significant impact on the final accuracy. \citep{reiner2005multi} As CNN increases its applicability and importance in the medical imaging domain, radiology workflows that enable AI models to access medical data are critically important. \citep{reiner2005multi, meenatchi2018survey}

Our design avoids the limitation of usability in the real world by adapting to radiology workflow and adhering to industry norms and standards. Although we did not evaluate the added value of the CADe to the diagnostic performance of radiologists, given the results of previous studies in which radiologists' diagnostic performance improved with deep learning algorithms, we believe that this algorithm can improve radiologists' performance even in large-scale population screening.

\section{Authorship statement}
All persons who meet authorship criteria are listed as authors, and all authors certify that they have participated sufficiently in the work to take public responsibility for the content, including participation in the concept, design, analysis, writing, or revision of the manuscript. Furthermore, each author certifies that this material or similar material has not been and will not be submitted to or published in any other publication.

\section{Ethical procedure}
The authors hereby declare that this research article meets all applicable standards with regards to the ethics of experimentation and research integrity. The authors also declare that the text of the article complies with ethical standards, the anonymity of the patients was respected.

\clearpage  
\addcontentsline{toc}{chapter}{Bibliography} 

\bibliographystyle{agsm}
\bibliography{main.bib}

\end{document}